\documentclass[aps,nofootinbib]{revtex4}
\newcommand{\dslash}{D\!\!\!\!\slash}
 \usepackage{amsmath}
 \usepackage{graphicx}
\usepackage{relsize}
\usepackage{enumitem}
\begin{document}

\title {The Silver Blaze Problem in QCD}

\author{Thomas D. Cohen}
\email{cohen@umd.edu}

\affiliation{Department of Physics, University of Maryland,
College Park, MD 20742-4111}

\begin{abstract}
This article provides a pedagogical introduction to the Silver Blaze problem. This problem refers to the difficulty of reconciling to perspectives on QCD with a chemical potential.  The first is the phenomenological fact that at $T=0$ QCD  remains in its ground state---the vacuum---with all physical observables unchanged whenever the magnitude of a chemical potential is less than some critical value.  The second is the fact that in functional integral treatments, the inclusion of any nonzero chemical potential changes all eigenvalues of the Dirac operator for every gauge configuration, leading to a natural expectation that the functional determinants also changes, which leads to the expectation that physical observables should be altered. The problem amounts to explaining why nothing happens below the critical chemical potential. By focusing on the eigenvalues of $\gamma_0$ times the Dirac operator rather than the Dirac operator itself, it is possible to show that for QCD with two flavors and identical quark masses, an isospin chemical potential with a magnitude less than $m_\pi$ (and no baryon chemical potential), or a baryon chemical potential of less than $\frac{3}{2} m_\pi$ (and no isospin chemical potential), the functional integerals at $T=0$ themselves remain unchanged in all configurations that contribute to the functional integral with non-vanishing weight. However, for $\mu_{\rm crit}\mu_B > \frac{3}{2} m_\pi$, the Silver Blaze phenomenon arises due to functional determinants having nontrivial phases that lead to cancellations between different gauge configurations. The mechanism leading to such cancellations remains unknown.
\end{abstract}
\maketitle
\section{Introduction}

One of the most famous exchanges in Arthur Conan Doyle's Sherlock Holmes stories occurs in the story {\it Silver Blaze}\cite{doyle1894memoirs}, which concerns a missing racehorse and a dead body. Inspector Gregory inquires, ``Is there any point to which you would wish to draw my attention?''

``To the curious incident of the dog in the nighttime.''

``The dog did nothing in the nighttime.''

``That was the curious incident,” remarked Sherlock Holmes.

The fact that the dog does nothing turns out to provide an essential clue. The relevant point for QCD and nuclear physics is that the fact that nothing happens can be highly significant, and one hopes it might ultimately give an essential clue about an important problem. The issue concerns the behavior of nuclear matter.

Nearly a century ago, von Weizsäcker\cite{vW}, in developing his semi-empirical mass formula, introduced the notion of cold infinite nuclear matter. Since then, the properties of infinite nuclear matter have been at the heart of nuclear physics. From a modern perspective, one would like to understand cold nuclear matter directly from QCD. There is a particularly important variant of this problem that occurs in nuclear astrophysics; it concerns the properties of neutron stars. Ideally, the properties of the cold matter equation of state (EOS) of electrically neutral matter would be computed directly from QCD; this is a critical ingredient in calculations of neutron star properties. Unfortunately, it is poorly constrained experimentally~\cite{NS1,NS2}. If it were possible to determine the cold matter EOS from QCD calculations, it would greatly aid neutron star physics. However, given the present state of the art, calculations of the cold matter equation of state from QCD are not viable.

There is a widespread consensus in much of the nuclear community that the only viable approach to numerically compute quantities in QCD, with errors under control systematically, is via Monte Carlo treatments of the Euclidean-space functional integral of a lattice version of the theory\footnote{It should be noted that alternative approaches, have been proposed.  One of these based on a truncation of the QCD Hamiltonian on the light cone (for a review, see \cite{Brodsky:1997de}), has not been developed to the point that they are competitive with lattice methods. A better developed class of approaches goes under the moniker of ``functional methods'' \cite{PhysRevD.102.034027} and includes such approaches as truncated Schwinger-Dyson equations and the functional renormalization group. It is unclear the extent to which these approaches will become competitive with lattice Monte Carlo calculations for {\it a priori} calculations of QCD observables. In any case, the discussion in this article is limited to  treatments of the Euclidean functional integral.}.

However, calculations of the EOS using conventional lattice methods are not viable; the inclusion of the chemical potential leads to an intractable sign problem \cite{SP0,SP1,SP2}.

The Silver Blaze problem\cite{Cohen:2003kd,Cohen:2004qp} is the difficulty of reconciling the functional integral formulation with the phenomenological fact that below some critical value for the absolute value of the chemical potential, cold QCD remains in its vacuum state. For a baryon chemical potential, the critical chemical potential is given by
\begin{equation}
\mu_{\rm cr} = M_{\rm nucleon} - B,
\end{equation}
where $M_{\rm nucleon}$ is the average of the proton and neutron masses, and $B$ is the binding energy of nuclear matter at saturation density—approximately 16 MeV. The difficulty arises because in functional integral formulations, the chemical potential enters only through the functional determinant. For simplicity, let us first consider the case of QCD with two flavors with identical quark masses.

The natural way to formulate the functional integral is to work at non-zero but low temperature in a large, but finite box of volume $V$, and then consider the behavior in the limit $T \rightarrow 0$ and the thermodynamic limit ($V \rightarrow \infty$) later. For the present purpose, it is useful to consider taking the limit $T \rightarrow 0$ before the limit $V \rightarrow \infty$; this will ensure that the physical spectrum of the theory is discrete throughout any calculations. The partition function for this system is given by a Euclidean space functional integral evaluated over a four-dimensional box with a temporal length of $\beta$ and periodic boundary conditions for the gluons:
\begin{equation}
Z_B(T,\mu_B) = \int d [A] e^{-S_{YM}} {\rm Det} \left( \dslash + m - \frac{{\mu_B\gamma_0}}{3}\right )^2 \; \label{funcint}
\end{equation}
where $S_{YM}$ is the Yang-Mills action. The physics of the quarks is captured by the functional determinant, which is computed using anti-periodic boundary conditions over the same box; in the functional determinant, $m$ is the quark mass, and $\mu_B$ is the baryon chemical potential (the factor of 1/3 reflects the fact that the baryon number of a quark is 1/3). Since we are considering a simplified theory with only two flavors—and they are degenerate—it is sufficient to compute the determinant for a single flavor and square it.

The presence of the chemical potential is manifest solely via the functional determinant, which is given simply as the product of the eigenvalues of the Dirac operator:
\begin{equation}
{\rm det}\left( \dslash + m - \frac{\mu_I \gamma_0}{2} \right) = \prod_j \lambda_j \; \; \; {\rm where} \; \; \; \left( \dslash + m - \frac{\mu_B \gamma_0}{3} \right) \psi_j = \lambda_j \psi_j.
\end{equation}

At this stage, a puzzle emerges: {\it A priori}, as soon as any chemical potential is introduced, in each gauge configuration every single eigenvalue of the Dirac operator differs from its $\mu_B=0$ analog. In the absence of a conspiracy of some sort among the eigenvalues, if every eigenvalue changes one expects that the functional determinant should change—and to change for every gauge configuration. And in the absence of some sort of conspiracy between gauge configurations, if every functional determinant changes then the natural expectation is that the partition function also changes. Yet at zero temperature, if we increase $\mu_B$ from zero, provided that $\mu_B < \mu_{\rm crit}$, the system remains in its vacuum state—and the partition function remains unchanged. That is, one increases $\mu_B$ and nothing happens—this is the analog of the curious incident of the dog in the nighttime. The Silver Blaze problem is how to reconcile the phenomenology of nothing happening with the functional integral treatment, in which all eigenvalues change in the functional determinant\cite{Cohen:2003kd,Cohen:2004qp}.

Before discussing the Silver Blaze problem in any detail, it is worth asking why it matters at all. Literally, nothing happens, so one might be tempted to dismiss the problem by suggesting that making a big deal about the Silver Blaze problem is literally {\it Much Ado About Nothing}. The argument against this temptation is simply that if one cannot understand why nothing happens when $\mu_B < \mu_{\rm cr}$, it is implausible that one could understand what is happening when a nontrivial ground state away from the vacuum forms. For example, if in the context of the functional integral approach, one could identify some criteria that must be satisfied to be in the Silver Blaze regime and develop a way to test the criterion in numerical calculations, one would have a straightforward way to calculate the binding energy of saturated nuclear matter. One key thing that should make the Silver Blaze situation simpler to understand than the generic case of finite density matter is that the state of the system is known—it is the vacuum. Apart from any potential future utility for the development of practical techniques to compute finite density QCD observables, solving the Silver Blaze problem would also be important from a purely theoretical perspective: it would help clarify how quantum field theories work in contexts that to date have not been well studied.

Before proceeding, it is also useful to specify what a solution to the Silver Blaze problem entails. The issue is less about understanding {\it why} the functional integral remains unchanged in the $T=0$ limit than about understanding {\it how} it remains unchanged. An analogy is useful here: the Banks-Casher formula\cite{Banks:1979yr}, which relates the chiral condensate to the expectation value of the density of states of the Dirac operator at zero virtuality. The Banks-Casher formula does not really explain why a non-zero chiral condensate forms but rather provides a mechanism for how it forms. In seeking a solution to the Silver Blaze problem, one wishes to find an analogous mechanism for how the system remains in the vacuum.

\section{The Silver Blaze phenomenon versus the Silver Blaze problem}

At this juncture, it is useful to distinguish ``the Silver Blaze problem,'' which arises in the context of functional integral treatments and concerns functional determinants, from something more general: ``the Silver Blaze phenomenon'' or ``Silver Blaze property.'' The phenomenon or property is simply the fact that at zero temperature, there is a critical value of the absolute value of the chemical potential below which the system remains in its ground state.

Over the years, many methods have been proposed to deal with sign problems in functional integrals—either by evading the sign problem entirely or by mitigating it to the degree that it can be overpowered numerically. For some recent reviews, see refs.~\cite{SPsol,SPsol0,Spsol1,SpSOl2}. However, so far, none of these methods are sufficiently powerful to render viable lattice treatments of QCD at nonzero chemical potential. Perhaps sometime in the future, quantum computers might provide a way forward. Unfortunately, the study of quantum field theories on quantum computers is still in the very early stages \cite{QC1,QC2,QC3,QC4,QC5,QC6,QC7,QC8,QC9}. Moreover, quantum computers with a very large number of error-corrected qubits, a necessary condition for QCD calculations, will not be available for a considerable time—if ever.

As noted above, the Silver Blaze problem can be thought of as a potential clue as to how the sign problem might be attacked fruitfully for cold matter. Regardless of this, the more general Silver Blaze phenomenon is a useful benchmark for any computational approach that aims to study the regime of low temperatures and low but non-zero chemical potentials. Note that if a computational approach were developed that works in this regime for some class of theory, it is not always necessary to understand {\it why} it works or, in a technical sense, how it works—{\it i.e.} to solve the Silver Blaze problem. It may often be sufficient to verify that it works. Ensuring that the system remains in its ground state below some critical value can be a very useful test of the viability of a method. Such benchmarking as a test of proposed methods, either for relatively simple theories with chemical potentials or with simplified approximations, has been used for more than 20 years (for some early examples see \cite{Adams:2004yy,Osborn:2005aj,Ravagli:2007rw}) and remains an active field of research (for some recent examples, see \cite{Balassa:2025bgt,Stoll:2025jor,Geissel:2025oyc,Topfel:2024iop,Pai:2024tip,Pannullo:2024sov,Iida:2024irv,Hidaka:2024drb,Podo:2023ute}). Indeed, to the extent that issues are associated with the development of the field, the Silver Blaze phenomenon as a tool for benchmarking has probably played a more central role than the Silver Blaze problem.

 \section{Two regimes of $\mu_B$  }

 Returning the Silver Blaze problems itself, 
 as was noted in the introduction, the failure of the naive expectation  that in functional integral treatments the partition function should change with changing $\mu_B$ even when $\mu_B < \mu_{\rm crit}$ requires some sort of conspiracy.   Either there is a conspiracy among eigenvalues of the Dirac operator to leave the functional determinant unchanged for configurations that contribute with nonzero weight to the functional integral, or the functional determinants of the relevant configurations change, but there is  a conspiracy among gauge configuration with the phases of the functional determinants leading to cancellations that precisely compensate for the effects of changed  magnitude of determinants.    Clearly, a key questions is which type of conspiracy is at work?  Remarkably the answer is both---depending on the regime.   
 
 In particular, for theories that have two flavors with degenerate masses, if
 \begin{equation}
|\mu_B| < \frac{3 m_\pi}{2} \, ,
 \end{equation}
 it can be shown that the conspiracy must be that as  $T \rightarrow 0$ the functional determinants do not change for configurations that contribute with nonzero weight---even though all of the eigenvalues making up the determinants change.  It turns out that it is trivial to show that this must be the case, at least for a theory with 2 flavors that are degenerate.  Moreover, it is comparatively easy to see how this can  comes about: {\it i.e.} there is a highly plausible simple solution to the Silver Blaze problem in this regime.  Moreover, it is plausible that the general behavior that the  nature of the conspiracy is such that the functional determinants of the relevant configurations are unchanged despite changed eigenvalues remains true whenever $|\mu_B|$ is less than some non-zero value  even when the quark masses are not degenerate.  
 
 However  the regime 
  \begin{equation}
 \frac{3 m_\pi}{2} < \mu_B <  \mu_{\rm cr}
 \end{equation}
the Silver Blaze  problems remains essentially unsolved, even for the case of degenerate quark masses.  It necessarily involves a conspiracy in which the functional determinants of the relevant configurations change with changing $\mu$, but cancellations between various gauge configuration with differing functional determinants act to leave the partition function unchanged.  Finding a solution in this regime remains an important challenge going forward.

 \section{The nature of the regime $\mu_B <  \frac{3 m_\pi}{2}$  \label{small}}
 
If one considers the case of QCD with two degenerate flavors it is easy to see why the conspiracy causing the Silver Blaze phenomenon must be that as  $T \rightarrow 0$,   the functional determinants will not change for all of the relevant configurations rather than being due to cancellations between the various gauge configurations.

This can be seen from a QCD inequality; for A review of QCD inequalities see \cite{Nussinov:1999sx}.  The QCD inequality of relevance, \cite{Cohen:2003ut} relates the partition function for  QCD  with two degenerate flavors with a baryon chemical potential to  the same system with an isospin chemical potential; it follows from the structure of the QCD functional determinant.   In QCD with an isospin chemical potential the up and down quarks each have opposite chemical potential each with a magnitude of $\frac{\mu_I}{2}$, while for  QCD with a baryon chemical potential  the up and down quarks have the same chemical potential of $\frac{\mu_B}{3}$.  Moreover, it is easy to show \cite{Cohen:2003ut}  that for a every quark flavor in a gauge configuration $A$, ${\rm det}_A$ including a chemical potential $\mu$ has the property that 
\begin{equation}
{\rm det}_A (\mu) = ({\rm det}_A (-\mu))^* \; .
\label{eq:Detcond}\end{equation}

Thus, the functional integrals for the partition function for the isospin and baryon cases are given by
\begin{equation}
\begin{split}
Z_I(\mu_I) &= \int [d A] e^{-S^{YM}_A } |{\rm det}_A(\mu_I/2)|^2 \;, \\ 
Z_B(\mu_B) &= \int [d A] e^{-S^{YM}_A } {\rm det}_A^2(\mu_B/3)  =\int [d A] e^{-S^{YM}_A}  |{\rm det}_A(\mu_B/3)|^2 e^2i \theta_A\\&= \int [d A] e^{-S^{YM}_A}  |{\rm det}_A(\mu_B/3)|^2 \cos(2  \theta_A)
\end{split} \label{Eq:ZB}
\end{equation}
where the integral is over all gauge configurations, $S^{YM}_A $ and ${\rm det}_A$ are the Yang-Mills action  and functional determinant for the configuration; implicitly this is done at fixed temperature which is imposed via standard thermal boundary conditions---periodic for gluons  (with period $\beta=1/T$) and antiperiodic for quarks.  The preceding expression is valid in when up and down quarks have the same mass.  The functional determinant can always be written as ${\rm det}_A(\mu)=|{\rm det}_A(\mu)| e^{i \delta_A(\mu)}$ and since the partition function is ultimately real  the imaginary part can be dropped and  the complex exponential replaced with a cosine.  

It is useful to rewrite these expressions in terms of the ratio the partition function with a chemical potential to the partition function at zero chemical potential
\begin{equation}
\begin{split}
\frac{Z_I(\mu_I)}{Z_0} & = \left  \langle \frac{|{\rm det}_A(\mu_I/2)|^2}{{\rm det}_A^2} \right \rangle \\ 
\frac{Z_B(\mu_B)}{Z_0} & = \left  \langle \frac{({\rm det}_A(\mu_B/3))^2}{{\rm det}_A^2} \right \rangle = \left  \langle \frac{|{\rm det}_A(\mu_B/3)|^2 \cos(2  \theta_A)}{{\rm det}_A^2} \right \rangle
\end{split}
\end{equation}
where the brackets indicate averaging over the gauge configurations weighted by a probability distribution given by $ P(A) \equiv e^{-S^{YM}_A } {\rm det}_A^2/ \left ( \int d[A] e^{-S^{YM}_A } {\rm det}_A^2\right )$ where ${\rm det}_A$ is the functional determinant in the absence of a chemical potential and is manifestly real.  The key point is that $\cos(2  \theta_A) < 1$ for all $\theta_A$ while the probability distribution is positive.  Thus at any temperature
\begin{equation}
    Z_I \left (\frac{2\mu_B}{3} \right) \ge  Z_B \left (\mu_B \right) \; , \label{Eq:ineq}
\end{equation}
which implies that the free energy for the isospin system is lower than for the baryon system.  

As it happens, there is also a Silver Blaze problem for an isospin chemical potential: phenomenologically, whenever the isospin chemical potential is less than $m_\pi$, the system remains in the vacuum state at zero temperature and the free energy density relative to the vacuum is necessarily zero. Moreover, since the partition function for the isospin only depends on the magnitude of $\left({\rm det}_A[\mu_I/2] \right)^2$ but not its phase, the isospin Silver Blaze cannot come about due to cancellations due to phases. Thus one is driven to the conclusion that the isospin Silver Blaze problem requires that the magnitude of the functional determinant for the configurations that have non-vanishing weight in the functional integral must be unchanged from the $\mu=0$ values. This in turn implies that for the case of the baryon chemical potential both the magnitude and phase of the functional integral are unchanged for configurations of non-vanishing weight in the functional integral: had the phases differed from zero with non-vanishing weight with the magnitudes fixed, the partition function would be smaller than in the vacuum, implying a free energy greater than the vacuum, which is not possible thermodynamically. The inequality $Z_I \left (\frac{2\mu_B}{3} \right) \ge  Z_B \left (\mu_B \right)$, in turn, implies that if the isospin Silver Blaze problem is solved, so is the baryon Silver Blaze problem---at least for the case where $m_u=m_d$ and $|\mu_B| < \frac32 m_\pi$.

\section{Properties of the functional determinant}

To proceed further, one needs to know some properties of the functional determinant when a chemical potential is included. Clearly the behavior of the system depends on the properties of the functional determinant of the Euclidean Dirac operator ${\rm det}_A(\mu)$---or more precisely the ratio ${\rm det}_A(\mu)/{\rm det}_A$. For a single flavor of quark this is given by
\begin{equation}
\frac{{\rm det}_A(\mu)}{{\rm det}_A} = \frac{Det\left [ D_\mu \gamma_\mu + m - \mu \gamma_0  \right ] }{Det\left [D_\mu \gamma_\mu + m  \right ] } = \frac{Det \left [\gamma_0 \left (D_\mu \gamma_\mu + m - \mu \gamma_0 \right )\right ] }{Det\left [\gamma_0 \left (D_\mu \gamma_\mu + m \right ) \right ] }  = \frac{Det \left [ D_0 + \sum_i D_i \gamma_0 \gamma_i  + m \gamma_0 - \mu \right ] }{ Det \left [  D_0 + \sum_i D_i \gamma_0 \gamma_i  + m \gamma_0  \right ] } 
\end{equation}
where $D$ indicates a covariant derivative; the second form follows from the fact that the determinant of a product is equal to the product of determinants and the index  
$i$ in the final form represents spatial directions.
This form is useful in that the chemical potential term leaves eigenfunctions of the operator $ D_0 + \sum_i D_i \gamma_0 \gamma_i  + m \gamma_0 $ unchanged when a chemical potential is included and is simply a constant added to the eigenvalues. Since the determinant is the product of the eigenvalues, $ \frac{{\rm det}_A(\mu)}{{\rm det}_A(0)} = \frac{\prod_k \left(\epsilon_k^R + i \epsilon_k^I - \mu \right )}{\prod_k \left(\epsilon_k^R + i \epsilon_k^I \right )} $ where $k$ labels the eigenfunctions and $\epsilon_k^R$ and $\epsilon_k^I$ are the real and imaginary parts of the eigenvalues. The eigenfunctions must satisfy the thermal boundary conditions for fermions: $\psi_k(t+ \beta)=-\psi_k(t+ \beta)$.

Moreover, Floquet's theorem\cite{Floquet1883} allows for a useful decomposition of the eigenvalues. The form of the operator $ D_0 + \sum_i D_i \gamma_0 \gamma_i  + m \gamma_0$, where any time dependence in the gauge fields is periodic in time, implies that if $\psi_k(t)$ is an eigenfunction satisfying the boundary conditions with eigenvalue $\epsilon_k^R + i \epsilon_k^I $ then $e^{i 2 \pi/\beta} \psi_k(t)$ is also an eigenfunction satisfying the thermal boundary condition; it has eigenvalue $\epsilon_k^R + i \epsilon_k^I + 2 \pi i$. Thus it is useful to consider the label $k$ as having two sublabels $j$ and $n$ and to write the eigenfunction $\psi_k(t)$ as $\phi_j(t) \exp (i (n+1) \pi t/\beta)$, where $\phi_j(t)$ is an ``intrinsic'' eigenfunction that is periodic $\phi_j(t)=\phi_j(t+ \beta)$ and an integer label $n$ for an antiperiodic exponential that generates eigenfunctions starting with the intrinsic form; eigenvalues can be written as $\epsilon_k^R + i \epsilon_k^I = \epsilon_j + \theta_j/\beta + \pi (n+1)/\beta$, where $\epsilon_j$ is real.
It is useful to regard $\epsilon_j$ as a ``quasi-energy''.

This decomposition into an intrinsic part times a complex exponential is not unique unless one imposes a condition on the allowable values of $\theta_j$, restricting its values to a range of $2 \pi$. Here the natural choice of $-\pi \le \theta_j < \pi$ will be taken, rendering the decomposition well defined.

In addition the operator $ D_0 + \sum_i D_i \gamma_0 \gamma_i  + m \gamma_0$ has a Hermitian part, $\sum_i D_i \gamma_0 \gamma_i  + m \gamma_0$ that anti-commutes with $\gamma_0 \gamma_5$ and an anti-Hermitian part, $ D_0 $, which commutes. This implies that if $\phi_{j_+}(t) \exp (i (n+1) \pi t/\beta)$ is an eigenfunction with eigenvalue $\epsilon_{j_+} + i \theta_{j_+}/\beta + i (2n+1) \pi/\beta$ (with $\epsilon_{j_+}>0$), then $\gamma_0 \gamma_5 \phi_{j_+}(t) \exp (i (n+1) \pi t/\beta)$ is also an eigenfunction; its eigenvalues are $-\epsilon_{j_+} + i \theta_{j_+}/\beta + i (2n+1) \pi/\beta$. The quasi-energies always come in pairs that are equal and opposite while the phases $\theta_j$ for the pairs are equal.

Given these properties of the eigenvalues the ratio of determinants can be expressed formally as
\begin{equation}
\begin{split}
  \frac{{\rm det}_A(\mu)}{{\rm det}_A}  &= \prod_{j_+,n,n'}  \frac{ \left((\epsilon_{j_+} - \mu)+ i \theta_{j_+}/\beta - (2n+1)  \pi / \beta  \right )\left((-\epsilon_{j_+} - \mu)+ i \theta_{j_+}/\beta - (2n'+1)  \pi / \beta  \right )}{  \left((\epsilon_{j_+} )+ i \theta_{j_+}/\beta - (2n+1)  \pi / \beta  \right )\left((-\epsilon_{j_+})+ i \theta_{j_+}/\beta - (2n'+1)  \pi / \beta  \right )} \\
  &= \Delta_A^+(\mu) \Delta_A^-(\mu) \; \;\;  \; {\rm with} \; \; \;  \; \Delta_{\pm}(\mu) = \prod_{j_+,n}  \frac{ \left((\pm \epsilon_{j_+} - \mu)+ i \theta_{j_+}/\beta - (2n+1)  \pi / \beta  \right )}{  \left((\pm \epsilon_{j_+} )+ i \theta_{j_+}/\beta - (2n+1)  \pi / \beta  \right )} . \label{eq:prodform}   
\end{split}
\end{equation}
where the product over $j_+$ only includes the positive quasi-energies.

As written Eq.~(\ref{eq:prodform}) needs to be regarded as formal since the product over $n$ in $\Delta_{\pm}$ does not converge. Fortunately, there is a natural way to regulate the expression. Note that if one differentiates the logarithm of the expression for $\Delta_{\pm} (\mu)$ with respect to $\mu$ twice, the resulting sum over $n$ is convergent. Integrating back yields a regularized expression, with the constants of integration determining the regularization. Clearly, one of these conditions is simply that $\Delta_{\pm}(0)=1$. The other integration constant can be deduced on physical grounds: the appropriate integration constant enforces the condition that $\lim_{\mu \rightarrow - \infty} \Delta_A^+(\mu)=1$; modes with positive quasi-energies should be interpreted as being largely unfilled particle states and in the limit of large negative chemical potentials they should be completely unfilled. Together these two conditions imply
\begin{equation}
\Delta_A^+(\mu) =\prod_{j_+} \frac{ 1+ \exp(-\beta (\epsilon_{j_+} -\mu)) e^{i\theta_{j_+}}}{1+ \exp(-\beta \epsilon_{j_+}) e^{i\theta_{j_+}}} \; \; .
\end{equation}
$\Delta_A^-(\mu)$ can then be fixed from the condition that ${\rm det}_A(\mu)=\left( {\rm det}_A(-\mu) \right)^*$ given in Eq.~(\ref{eq:Detcond}): $\Delta_A^-(\mu) =\left(\Delta_A^+(-\mu)\right )^*$.
In effect $\Delta_A^+(\mu)$ takes into account the effects of adding particles while $\Delta_A^-(\mu)$ is the analogous effect for producing holes---i.e. antiparticles.

Our interest is for very low temperatures. For $\mu >0$, it should be apparent that as $\beta \rightarrow \infty$, the factors in the product over $j_{+}$ in $\Delta_A^{-}$ all tend toward unity, as do the factors with $\epsilon_{j_+} < \mu$ in $\Delta_A^+$. On the other hand, the contributions with $\epsilon_{j_+} < \mu$ in $\Delta_A^+$ become exponentially large and dominate; there is analogous behavior for $\mu$ negative. So that as the low temperature limit is approached
\begin{equation}
\begin{split}
\frac{{\rm det}_A(\mu)}{{\rm det}_A }&= \exp \left (\sum_{j_+}\Theta(|\mu|-\epsilon_{j_+}) \left (  \beta (\mu-\epsilon_{j_+} ) + i \theta_{j_+} {\rm sgn}(\mu)  \right  ) \right)\times \left ( 1 + {\cal O}\left (\exp \left (-\beta (\mu-\epsilon_{j_+^{\rm min}} )\right )  \right)  \right )\\ &\xrightarrow[\beta \rightarrow \infty ]{\text{}} \:\left ( \frac{{\rm det}_A(\mu)}{{\rm det}_A}  \right)_{\rm asym} \!\!
\!=\exp \left (\sum_{j_+}\Theta(|\mu|-\epsilon_{j_+}) \left (  \beta (\mu-\epsilon_{j_+} ) + i \theta_{j_+} {\rm sgn}(\mu)  \right  ) \right) \;
\end{split} \label{Eq:asym}
\end{equation}
where $\Theta$ is the Heaviside step function, which is unity for positive arguments and zero for negative arguments, ${\rm sgn}$ is the sign function which is +1 for positive arguments and -1 for negative arguments and $\epsilon_{j_+^{\rm min}}$ is the smallest value of $\epsilon_{j_+}$ in the spectrum for a given gauge configuration.

It is convenient to rewrite the expression for the asymptotic form  in Eq.~(\ref{Eq:asym}) in terms if the following two quantities:
\begin{equation} 
\begin{split}
    N_A(\mu) &= \sum_{j_+} \Theta (|\mu| - \epsilon_{j_+}) \; \; \; {\rm and} \; \; \; \delta_A(\mu) = \sum_{j_+} \Theta (|\mu| - \epsilon_{j_+}) \theta_{j_+} \, \\
\left ( \frac{{\rm det}_A(\mu)}{{\rm det}_A}  \right)_{\rm asym} \!\!\! &= \exp \left( \int_{0}^{|\mu|} \, d \nu \, \left (\beta N_A'(\nu)  (|\mu| -\nu)\right) \right)\exp \left(   i \int_{0}^{|\mu|} \, d \nu  \, \delta_A'(\nu)\right)  \; .\label{Eq:asym2}
\end{split}
\end{equation}

A few comments are in order. First, note that in the zero temperature limit, the step function in Eq.~(\ref{Eq:asym}) ensures that 
the functional determinant is unchanged from its $\mu=0$ value whenever $|\mu|$ is less than the smallest positive quasi-energy, $ \lim_{T \rightarrow 0} \frac{{\rm det}_A(\mu)}{{\rm det}_A}  = 1 $.
This plays an essential role in solving the Silver Blaze problem---at least in certain regimes. Moreover, this comes about despite the fact that every eigenvalue of the Dirac operator, and as is relevant here, every eigenvalue of $\gamma_0$
times the Dirac operator is also altered---they all shift by $\mu$. 
Second, the leading correction to the asymptotic form is exponentially suppressed at low temperatures by a factor of relative order ${\cal O}\left (\exp \left (-\beta (|\mu|-\epsilon_{j_+^{\rm min}} )\right )  \right)$. 
Third, note that in 
the products in Eqs.~(\ref{eq:prodform}) and the sum in Eq.~(\ref{Eq:asym}) it is assumed that the spectrum of $\epsilon_{j_+}$ is discrete rather than continuous. If the spatial volume is large but finite one expects the spectrum to be discrete. This is important as it ensures that for sufficiently low temperatures the dominant error will be associated with the error in using a step function (rather than the exponential form) for the eigenvalue with the lowest value of $|\epsilon|$.

A final comment: it is important to understand the nature of the low temperature limit in Eq.~(\ref{Eq:asym}). The limit should be seen as a mathematical one in which the temperature is low enough so that the difference between the exponential and step function is negligible. However, this limiting form does not directly correspond to lowering the temperature of a physical system since in addition to forcing the exponential form to be close to the step function, lowering the temperature alters the boundary conditions which alters the eigenvalues.

\section{A solution to the isospin Silver Blaze problem for 2-flavor QCD with degenerate quark masses \label{Sec:Iso} }

The case of two degenerate flavors is considered here as it is particularly straightforward. With that restriction the asymptotic form in Eq.~(\ref{Eq:asym}) suggests a very natural solution to the isospin Silver Blaze problem---and with it the solution to the baryon Silver Blaze problem in the regime $|\mu_B| < \frac{3 m_\pi}{2}$. This section works through in considerable detail the mathematical steps needed to show this, along with the phenomenologically inspired assumptions necessary to make it work. However at its heart, the solution amounts to the fact that $ \lim_{T \rightarrow 0} \frac{{\rm det}_A(\mu)}{{\rm det}_A}  = 1 $ along with the assumption that the spectrum for $\epsilon_{j_+}$ is gapped.

To see how this works, it is useful to note that while the form of the eigenfunctions of $ D_0 + \sum_i D_i \gamma_0 \gamma_i  + m \gamma_0$ obviously change under gauge transformations, the associated eigenvalues are invariant. Since they are invariant they can be regarded as being physical observables. Moreover, all of the information about the asymptotic spectra for any configuration is contained in the functions $N_A(\mu)$ and $\delta_A(\mu)$ defined in Eq.~(\ref{Eq:asym2}); moreover, in the zero temperature limit $N_A(\mu)$ has a very simple physical interpretation.

Recall that the partition function for a system at $T=0$ with an isospin chemical potential is given by $\frac{Z_I(\mu_I)}{Z_0 } = \left \langle \frac{|{\rm det}_A(\mu_I/2)|^2}{{\rm det}_A^2} \right \rangle$ where the averaging is over gauge configurations weighted by $e^{-S^{YM}_A } {\rm det}_A^2/ \left ( \int d[A] e^{-S^{YM}_A } {\rm det}_A^2\right )$. Inserting Eq.~(\ref{Eq:asym2}) into this form and taking the zero temperature limit yields 
\begin{equation}
 \frac{Z_I(\mu_I)}{Z_0 }= \left \langle \exp \left( \int_{0}^{\mu_I/2} \, d \zeta \, \left ( \beta N_A'(\zeta) (\mu_I -2 \zeta)\right) \right)\right \rangle_{\rm vac} = \left \langle \exp \left( \frac12 \int_{0}^{\mu_I} \, d \nu \, \left ( \beta N_A'(\nu/2) (\mu_I - \nu)\right) \right)\right \rangle_{\rm vac}
 \label{Eq:I3form1}
\end{equation}
where the $T=0$ limit gives the vacuum expectation values and $N_A$ is evaluated at $T=0$.

On the other hand, for any finite system, the thermodynamics at asymptotically low values of the temperature allows one to write $\frac{Z_I(\mu_I)}{Z_0 }$ in terms of $I_3(\mu_I)$, the value of the third component of the isospin at fixed isospin chemical potential: 
the energy of the vacuum state as a function of $\mu_I$ is given by $E(\mu_I) = \int d \nu \, \nu I_3'(\nu)$ (where $E$ is measured relative to the vacuum) so that
\begin{equation}
 \frac{Z_I(\mu_I)}{Z_0 } = \left \langle \exp\left (-\beta (E(\mu_I)- \mu_I I_3(\mu_I))\right ) \right \rangle_{\rm vac } = \left \langle \exp \left( \int_{0}^{\mu_I} \, d \nu \, \left ( \beta I_3'(\nu) (\mu_I - \nu)\right) \right) \right \rangle_{\rm vac} \, .\label{Eq:I3form2}
\end{equation}
Comparing Eqs.~(\ref{Eq:I3form1}) and (\ref{Eq:I3form2}), it is apparent that when the chemical potentials for the two species of quark are equal and opposite, the operator $N_A(\mu_I/2)$ evaluated in the vacuum state is identical to $I_3(\mu_I)$; one should interpret $N_A$ as simply being the third component of the isospin in this context. This equivalence must hold for the vacuum state generally and not just for $ \frac{Z_I(\mu_I)}{Z_0}$. In particular it implies that 
\begin{equation} 
\left \langle I_3(\mu_I) \right \rangle_{\rm vac } = \left \langle N_A(\mu_I/2) \right \rangle_{\rm vac } \; .
\end{equation}

This interpretation makes intuitive sense. In a theory of non-interacting up and down quarks (with degenerate masses) at zero temperature the isospin is simply the number of filled (positive energy) up quark single particle states with energies less than $\mu_I/2$. What the rather involved analysis up to this point demonstrates is that even when the particles interact via the exchange of gauge bosons, an analogous thing holds with $\epsilon_{j_+}$ replacing the single particle energies as the relevant quantity.

The key point is that apart from $N_A(\mu_I/2)$ being interpretable in terms of the isospin, it also has a direct interpretation in terms of the spectrum of $\gamma_0$ times the Dirac operator. The fact that the vacuum remains unchanged for $\mu_I \le m_\pi$ is simply equivalent to the existence of a gapped physical spectrum for $\epsilon_{j_+}$ at $T=0$. But the fact that a $T=0$ physical spectrum has a gap is a very ordinary occurrence and no great conspiracy is needed for it to occur: there is no need for remarkable cancellations.

That said, one interesting issue is worth noting. The physical spectrum is obtained by taking the appropriate weighted average over gauge configurations, each of which contributes with a positive weight. Consider a configuration in which the gauge field is exactly zero. Such a configuration surely exists and should be included in the weighted average. That configuration is gapped, but its gap is equal to the quark mass, $m$. Of course, the precise value of $m$ depends on how the theory is regularized, but typical regularizations lead to a quark mass of a few MeV---which is well below $m_\pi/2$, the physical gap. Moreover, Goldstone's theorem~\cite{Goldstone:1961,Goldstone:1962vj} implies that the spontaneous breaking of QCD's approximate chiral symmetry means that the pion is a pseudo-Goldstone boson. This leads directly to the Gell-Mann--Oakes--Renner relation~\cite{GellMann:1968rzm}, which has the pion mass scaling with the square root of the quark mass and implies that for sufficiently small quark masses, the quark mass is always qualitatively larger than the physical gap. Presumably, apart from the configuration with zero gauge field, there are numerous other gauge configurations that have gaps less than $m_\pi/2$. One might worry that since such configurations cannot have negative weight in the averaging, their effects cannot cancel out and that inevitably the physical gap would go down at least as far as $m$, which is incompatible with what is observed phenomenologically. However, such a concern is misplaced. While it is indeed true that configurations with gaps below $m_\pi/2$ do exist, there is no necessity that such configurations contribute to the vacuum with nonzero weight: the integrated contribution to the vacuum from all such configurations can be zero without any cancellations. All that is necessary is for such configurations to have zero measure in the appropriate averaging; this can easily occur given that the number of configurations that do contribute is infinite. Clearly, this is what occurs; moreover, the same mechanism that is responsible for spontaneous chiral symmetry breaking will also cause the gap to scale with the square root of the quark mass at small $m$.

\section{The baryon Silver Blaze problem for 2-flavor QCD with degenerate quark masses \label{Sec:Bary}}

As noted in Sec.~\ref{small}, when $\mu_B <  \frac{3 m_\pi}{2}$ for 2-flavor QCD with degenerate quark masses, the solution of the isospin Silver Blaze problem implies a solution of the baryon Silver Blaze problem.   This is due to the fact that for every configuration the  functional determinant in the baryon problem has the same magnitude as the in the analogous isospin problem but comes with a phase factor.  The effect of the phase factor is that on a configuration by configuration basis, the real part of the functional determinant is  always less that that in the isiospin case.  Thus if the system has the same partition function as the vacuum state, then so must the baryon case.  There is a related issue. Since the inclusion of a baryon chemical potential introduces a phase factor, and the isospin chemical potential keeps the system in its vacuum state when sufficiently small, something must prevent the phase factor from pushing the partition function below that of the vacuum. Clearly, vacuum stability implies on physical grounds that this cannot happen. However, it is instructive to understand at a mathematical level how the functional integral prevents this from happening. Fortunately, it is quite straightforward. If one looks at the definitions of $N_A(\mu)$ and $\delta_A(\mu)$ in Eq.~\eqref{Eq:asym2}, one sees that they both only get contributions when $\mu$ is greater than the smallest value of $\epsilon_{j_+}$. Thus when $\mu$ is inside the gap, $\theta_A$ is automatically zero, the phase factor becomes unity and does not lower $Z$.

While the Silver Blaze problem for the regime $\mu_B < \frac{3 m_\pi}{2}$ is basically well understood, an understanding of the regime $\mu_{\rm crit} > \mu_B > \frac{3 m_\pi}{2}$ remains elusive after more than two decades after the problem was introduced. The reason the problem is unsolved is that the Silver Blaze phenomenon implies that $\frac{Z_B(\mu_B)}{Z_0}$ at $T=0$ must be unity in this regime, but this can only come about if there is an intricate interplay between the expectation value of the magnitude and the phase of ${\rm det}_A(\mu_B)$.  The  mathematical origin of this interplay obscure.

In particular, from Eq.~\eqref{Eq:ZB} one can write  
\begin{equation}
\frac{Z_B(\mu_B)}{Z_0} = \frac{Z_I(\frac{2}{3} \mu_B)}{Z_0} \left \langle \cos(2 \theta_A)\right \rangle \; \; {\rm with} \; \;
\left \langle \cos(2 \theta_A)\right \rangle  = \frac{\int [d A] e^{-S^{\rm YM}_A}  |{\rm det}_A(\mu_B/3)|^2 \cos(2 \theta_A)}{\int [d A] e^{-S^{\rm YM}_A}  |{\rm det}_A(\mu_B/3)|^2 } \; ,
\end{equation}
\textit{i.e.}, $\left \langle \cos(2 \theta_A)\right \rangle$ is the average of $\cos(2 \theta_A)$ weighted by the probability in the analogous isospin matter state.

The remarkable thing is that for asymptotically small $T$, in order for $\frac{Z_B(\mu_B)}{Z_0}$ to approach unity---as it does phenomenologically whenever $\mu_{\rm cr}>\mu_B$---it must be true that  
\begin{equation}
   \left \langle \cos(2 \theta_A)\right \rangle = \frac{Z_0}{Z_I(\frac{2}{3}\mu_B)}=\frac{1}{\left \langle \exp \left( \int_{0}^{\frac{2}{3}\mu_B} \, d \nu \, \left ( \beta I_3'(\nu)  (\frac{2}{3}\mu_B - \nu)\right) \right) \right  \rangle_{\rm vac}} \; . \label{Eq:cond}
\end{equation}
In terms of the functions $N_A$ and $\delta_A$ defined in Eq.~\eqref{Eq:asym2}, this means
\begin{equation}
\left \langle \cos \left ( 2 \int_{0}^{\mu_B/3} \!\!\! d\nu \, \, \delta_A(\nu) \right )\right \rangle =\frac{1}{\left \langle \exp \left( \int_{0}^{\frac{2}{3}\mu_B} \, d \nu \, \left ( \beta N_A'(\nu/2)  (\frac{2}{3}\mu_B - \nu)\right) \right) \right  \rangle_{\rm vac}} \; . \label{Eq:cond2}
\end{equation}
That is, the phenomenology in the Silver Blaze regime requires a correlation between the expectation value of the phase factor that precisely cancels any growth in the magnitude of the determinant. The solution of the baryon Silver Blaze problem for the case of QCD with two degenerate quark masses amounts to explaining why Eqs.~\eqref{Eq:cond} and \eqref{Eq:cond2} should hold from the perspective of the functional integral without making an appeal to phenomenology.

\section{Open problems and future prospects}

 There are numerous open problems associated with Silver Blaze phenomena, beyond the baryon problem with two degenerate quark masses.  Some of these might easier to solve than others.  The solutions of any any of these may help to gain insights into others.  In this section the nature of a number of these will be outlined briefly and brief assessment of the prospects for progress will be made.  

 \subsection{The dependence of the gap on the quark mass for QCD 2-degenerate quark masses \label{Subsec:gap}}

 In Sec.~\ref{Sec:Iso} it was noted that the size of the gap in the spectrum of $\epsilon_{j_+}$ must scale as the square root of the quark mass (in the limit of small quark masses) due to chiral symmetry. It would be instructive to understand exactly how this comes about. Recall that the Gell-Mann--Oakes--Renner relation~\cite{GellMann:1968rzm} implies that the coefficient of proportionality is linear in the chiral condensate, which the Banks-Casher relation~\cite{Banks:1979yr} relates to the density of states of the Dirac operator at zero virtuality. It is quite plausible that an understanding of this issue could be achieved; doing so would presumably involve recasting the expression for the chiral condensate into a form that depends on the spectrum of $\gamma_0$ times the Dirac operator rather than the Dirac operator itself.

\subsection{Silver Blaze behavior with distinct chemical potentials for each flavor }

Sections \ref{Sec:Bary} and \ref{Sec:Iso} consider situations where the chemical potentials for the two species of quark were either the same or equal and opposite. However, there is a more general problem in which the chemical potentials for the two species are different in magnitude.  Let us initial restrict attention to cases with two flavors with the same quark mass but with distinct chemical potentials for the two species. In general the phenomenology in such cases can involve more than one transition. The phenomenology may be easier to think about as the system having an isospin chemical potential, $\mu_I = (\mu_u - \mu_d)$ and a baryon chemical potential $\mu_B = \frac{3}{2}(\mu_u + \mu_d)$.

Consider for example the situation where there is a chemical potential, $\mu_u$ for the up quark but a zero chemical potential for the down quark. Thus $\mu_I = \mu_u$ and $\mu_B = \frac{3}{2} \mu_u$. As one increases the up quark chemical potential from zero, phenomenologically at $T=0$ the system will remain in its ground state until $\mu_u = m_\pi$ at which point it becomes energetically favorable to form a state with a non-zero isospin density but zero baryon density. Indeed, the state is identical to one with $\mu_I = m_\pi$ and $\mu_B = 0$ as the baryon number is zero.  This occurs because even though there was no chemical potential for the down quarks, it becomes energetically favorable to produce down quarks and pair them with up quarks.

Now, it is easy to see that the same argument that was made for the isospin chemical potential (and for the baryon chemical potential below $3 m_\pi/2$) will explain the Silver Blaze phenomenon below $\mu_u = m_\pi/2$: the chemical potential lies below the gap. However, in the regime of $m_\pi > \mu_u > m_\pi/2$, phase cancellations similar in kind to those in the case of the baryon chemical potential must occur leaving the system in its vacuum state. It remains an open question as to how these occur. The interesting and important point is that while the phase cancellations are similar in kind to those in the baryon chemical potential case, they are not the same;  one might hope that the differences between the way the phase cancellations work in this context and in the baryon chemical potential context could contain clues for potential solutions to both Silver Blaze problems.

Note that if the up quark chemical potential continues to increase beyond $m_\pi$, the isospin density will continue to increase with zero baryon number. At some point, the up quark chemical potential becomes large enough to induce a phase transition to a regime with nonzero baryon density. Beyond the fact that the critical chemical potential for such a transition will be greater than three times the critical chemical potential for the baryon chemical potential with zero isospin chemical potential, at present  the value of the critical up chemical potential  for this to occurs is  unknown. While the system below this transition but above the transition to a state with nonzero isospin does not strictly represent a Silver Blaze phenomenon, as the system is not in the vacuum, it is somewhat analogous.

The more general case with distinct up and down quark chemical potentials can also provide information about the nature of phase cancellations. Without loss of generality, one can consider situations where the magnitude of the up quark is larger than the magnitude of the down quark. Phenomenologically the system will remain in the ground state when $\mu_I \equiv \mu_u - \mu_d$ is less than $m_\pi$ and $\mu_B = \frac{3}{2}(\mu_u + \mu_d)$ is below some critical value that will depend on $\mu_I$. From previous arguments if both $\mu_u$ and $\mu_d$ are both within the gap (\textit{i.e.}, if both $|\mu_u|$ and $|\mu_d|$ are less than $m_\pi/2$), then the Silver Blaze problem is solved without further work since the functional determinant will be unchanged from the vacuum. However when $\mu_u > m_\pi/2$, \textit{i.e.}, is outside the gap, but $\mu_d$ is inside the gap with $|\mu_d| < m_\pi/2$, there must be phase cancellation to preserve the Silver Blaze phenomenon, and in general these are distinct from the case $\mu_d = 0$ so again can provide additional information about how the phase cancellations work.

\subsection{Unequal quark masses}

The analysis in Sections \ref{Sec:Bary} and \ref{Sec:Iso} considered the case in which the up and down quark masses were degenerate. That situation had the virtue that in the case of an isospin chemical potential the phase factors for the up and down quark functional determinants canceled on a configuration-by-configuration basis, which automatically implied that there needed to be a gap in the spectrum for $\epsilon_{j_+}$. 

If the quark masses are different the phase factors no longer need to cancel perfectly and thus there is no formal guarantee that the $\epsilon_{j_+}$ spectra for the up and down quarks will be gapped. It remains an open question whether or not they are gapped. However it is quite plausible that the spectra are gapped.  If so, when the chemical potentials for the up and down quarks are both inside the gaps, then as before the Silver Blaze problem is solved. Of course, if one or both chemical potentials are outside the gap then phase cancellations are required for the Silver Blaze phenomenon to occur. Even if the spectra are gapped, the gaps for the up and down quarks will presumably occur at different places. How such gaps, if they exist, are related to $m_\pi$ remains an interesting open question.

\subsection{Future prospects}

What are the prospects of resolving the various open questions associated with Silver Blaze problems? It is helpful to recall a quip that Niels Bohr is reported~\cite{Mencher1971} to have made: ``Predictions are hard, particularly about the future.''\footnote{This notion has also been ascribed to Yogi Berra, the New York Yankee catcher, coach and part-time philosopher.} Given this general difficulty, assessing the prospects of progress on the various open issues associated with the Silver Blaze problem is not straightforward. That said, it seems plausible that the issue for which there is most likely to be progress in the near future is the one discussed in Subsection~\ref{Subsec:gap}. The issue there is how to understand the relationship between the spectral gap and the quark mass. It seems possible that this issue can be resolved because we already know how the pion mass scales with the square root of the quark mass and we know phenomenologically that the gap is fixed at $m_\pi/2$. Thus resolving this issue amounts to a formal understanding of how the gap opens to an amount given by $m_\pi/2$. While this issue might be understood in the not so distant future, it is also likely that such an understanding will not help resolve the baryon Silver Blaze problem in the regime $\mu_{\rm cr} > \mu_B > \frac{3}{2} m_\pi$; resolving this involves understanding subtle cancellations associated with phases.

Regarding the baryon Silver Blaze problem in that regime, it is worth recalling that there have been unsuccessful attempts to evade the sign problem for QCD with a chemical potential at low temperatures for several decades. Moreover, there has been no meaningful progress on the baryon Silver Blaze problem during the nearly quarter of a century since it was proposed. Thus, it is plausible that the issue will not be resolved in the near future. However, it remains possible that some progress can be made toward at least understanding the nature of the cancellations necessary to ensure the Silver Blaze phenomenon. A potential clue might be found by considering the variation with chemical potential for the needed cancellations for both the baryon Silver Blaze problem and the needed cancellations for the case of $\mu_u$ with $\mu_d = 0$. The first of these depends on the weighted average of $\cos(2\theta_A)$ while the second depends on the weighted average of $\cos(\theta_A)$. Thus they provide complementary information. Additional information might come from configurations related by center transformations---transformations that resemble gauge transformations but which have gauge potentials that are not periodic in $\beta$; the gauge potential at Euclidean time $\tau + \beta$ (where $\beta$ is the inverse temperature) is equal to an element of the group center times the gauge potential at $\tau$. Theses may give some insight since all of the quasi-energies, $\epsilon_{j_+}$, of two configurations connected by a nontrivial center transformation are the same, but the the phases differ by an element of the center. It remains to be seen whether significant insights can be obtained from such considerations.

 \begin{acknowledgments}

This work was supported in part by the U.S. Department of Energy, Office of Nuclear Physics under Award Number(s) DE-SC0021143, and DE-FG02-93ER40762.

\end{acknowledgments}

 \bibliography{main.bib}
\end{document}